\documentstyle[preprint,aps]{revtex}

\begin{document}
\draft

\title{Exact quantum state for $N=1$ supergravity}

\author{Andr\'as Csord\'as\thanks{Permanent address: Research Institute
for Solid State Physics, P.O. Box 49, H 1525 Budapest, Hungary} and
Robert Graham\\
Fachbereich Physik, Universit\"at-Gesamthochschule Essen\\
45117 Essen\\Germany}

\maketitle

\begin{abstract}
For $N=1$ supergravity in $3+1$ dimensions we determine the graded algebra
of the quantized Lorentz generators, supersymmetry generators, and
diffeo-morphism and Hamiltonian generators and find that, at least
formally,
it closes in the chosen operator ordering. Following our recent
conjecture and generalizing an ansatz for Bianchi-type models we proposed
earlier we find an explicit exact quantum solution of all constraints in
the metric representation.
\end{abstract}

\narrowtext

Supersymmetry has long been recognized not only as a possible answer to
some of the outstanding problems of particle physics and cosmology, but
also as a mathematical structure whose presence, in spite of its seemingly
higher complexity, can actually simplify field theories in an important
manner. Striking examples are the proof of the positivity of energy or
the recent proof of quark-confinement in such theories \cite{1a,1b}.
Another
example will occur in the present work, where we find that supersymmetry
helps to demonstrate formal closure of the generator algebra of the
constraints after quantization. In supergravity, like in pure gravity,
the study of spatially homogeneous models has recently provided some
insight, limited as it may be, in the non-perturbative canonical
quantization and in the physical states, in the sense of Dirac \cite{2},
solving the quantum
constraints. It was found recently that infinitely many physical states
exist even in such strongly reduced models \cite{3,4}. Even earlier some
states could be determined explicitely: For homogeneous supersymmetric
models of Bianchi-type in class A of Ellis and McCallum \cite{5}, in the
case where no matter fields beyond the Rarita-Schwinger field are present,
states in the empty and full fermion sectors could be found
[7-11] which have been interpreted as worm-hole states \cite{7}.
More physical states near the middle of the fermion-number spectrum,
between the empty and full fermion-number state, were found to
have the form \cite{3,4} $\Psi=S^\alpha S_\alpha
g(h_{pq})|0\rangle$, where $h_{pq}$ is the metric
tensor on the
space-like slices of space-time, $S_\alpha$, $\bar{S}_{\dot{\alpha}}$, are
the supersymmetry generators \cite{10} and $|0\rangle$ is the vacuum
which is annihilated by the
gravitino field, ${\psi_p}^\alpha|0\rangle=0$. One of these states could
again be
determined explicitely and has been identified as a Hartle-Hawking state
\cite{4}. So far all attempts \cite{11} to generalize the physical states
in the empty and full fermion sector of the Bianchi models to full
supergravity were unsuccessful, because it could be shown \cite{12,13,14}
that in full supergravity physical states in the zero fermion-number
or anti-fermion-number sectors, (which we shall call the bare vacua,
in the following), do not exist.

However, recently we made the conjecture \cite{3} that the
above mentioned form of the physical states about half-way between the
bare fermion and
anti-fermion vacua has a direct counterpart in full supergravity. It is
the purpose of the present paper to substantiate this conjecture by
formal calculations, which would, in fact, constitute a proof if they
were made rigorous by paying due attention to appropriate regularizations
and their subsequent removal when passing to the limit in the final result.

The explicit result we obtain in the sector half-way between the bare
fermion and anti-fermion vacua has an amplitude which reduces to the
worm-hole amplitude first found in \cite{6,7} for Bianchi-type IX and
extended in \cite{8a,8b,9} to the other Bianchi-types in class A. As
the latter states
reside in one of the two bare vacua they would seem to be indirect
spatially homogeneous counterparts of our general state, at best.
Nevertheless they turn out to be its {\it only} counterparts:
The direct spatially homogeneous restriction of our state including the
restriction to a spatially homogeneous Rarita-Schwinger field turns out
to vanish identically \cite{4} in the fermion sector half-way between
the bare vacua, even though we can show that the unrestricted state is
non-zero in general.

Let us begin now by first presenting our results on the graded algebra
of the constraint operators. We briefly recall the steps in the derivation
of the latter (see e.g. \cite{15,16}) and use this to fix our notation.
The starting point is the Lagrangean of $N=1$ supergravity \cite{10} in
the tetrad-representation from which the time-derivative of the
spin-connection is eliminated by adding an appropriate 3-surface term.
A (3+1)-decomposition is performed introducing an arbitrary foliation of
space-time by a continuous family of space-like 3-surfaces, which we shall
assume to be compact, to avoid surface terms, for simplicity. The basic
variables are then the tetrad fields ${e_i}^a(\bbox{x})$, and the
Rarita-Schwinger
field ${\psi_i}^\alpha(\bbox{x})$, ${\bar{\psi}_i}^{\ \dot{\alpha}}
(\bbox{x})$ where the space-like Einstein indices
$i=1,2,3$ are from the
middle and the Lorentz indices $a=0,1,2,3$;
$\alpha=1,2$; $\dot{\alpha}=1,2$ from the beginning of the alphabet.
$n^a$, a function of the ${e_i}^a$, will denote the future oriented normal
vector orthogonal on the space-like 3-surfaces, $n^an_a=-1$,
$n^ae_{ia}=0$.
Next, canonical momenta ${\hat{p}^i}_{\ a}(\bbox{x})$,
${\hat{\pi}^i}_{\ \alpha}(\bbox{x})$,
$\bar{\hat{\pi}}^i_{\ \dot{\alpha}}(\bbox{x})$ and associated Poisson
brackets are defined
as usual. This brings out the fact that certain additional constraints
exist in this theory: the Lorentz constraints $J_{\alpha\beta}\approx 0
\approx\bar{J}_{\dot{\alpha}\dot{\beta}}$ (where $\approx$ denotes weak
equality in the sense of Dirac \cite{2}), which turn out to be
first-class constraints; and second class constraints relating the
Grassmannian variables and their momenta. The second-class constraints are
duly eliminated by passing from Poisson brackets to Dirac brackets, and
the latter are simplified by introducing new non-canonical momenta
$\bar{\pi}^i_{\phantom{i}\dot\alpha}=
2{\bar{\hat{\pi}}^i}_{\phantom{i}\dot\alpha}=
-\varepsilon^{ijk} e_j^{\phantom{j}a} \psi_k^{\phantom{k}\alpha}
\sigma_{a\alpha \dot\alpha}$,
$P_{-\phantom{i} a}^{\phantom{-}i}=
\hat{P}^i_{\phantom{i} a}-\frac{1}{2}\varepsilon^{ijk}
C^{\dot\beta \alpha}_{lj} \sigma_{a \alpha \dot\alpha}
\bar\psi_k^{\phantom{k}\dot\alpha} \bar\pi^l_{\phantom{l} \dot\beta} $
with $C^{\dot\alpha \alpha}_{ij}=\frac{1}{2 h^{1/2}}
\left[ -i h_{ij} n^a + h^{1/2} \varepsilon_{ijk}e^{ka} \right]
\bar\sigma_a^{\phantom{a}\dot\alpha \alpha}$.
Passing to the Hamiltonian by the usual Legendre transformation and
adding the first-class constraint with its Lagrange multiplier we obtain
the total Hamiltonian as the usual sum of generators multiplied by their
Lagrange multipliers. After canonical quantization in the $({e_p}^a,
{\bar{\psi}_p}^{\ \dot{\alpha}})$-representation based on the Dirac
brackets where ${\bar{\pi}^i}_{\dot{\alpha}}=
-i\hbar\delta/\delta{\bar{\psi}_i}^{\ \dot{\alpha}}$,
$P_{-\phantom{i} a}^{\phantom{-}i}=-i\hbar\delta/\delta{e_i}^a$, the
constraint operators in a conveniently (but otherwise
arbitrarily) chosen operator ordering \cite{18a} become
\begin{eqnarray}
S_\alpha &=&{i\over2} P_{-\phantom{i}a}^{\phantom{-}i}
\sigma^a_{\phantom{a}\alpha \dot\alpha}
\bar\psi_i^{\phantom{i}\dot\alpha}+
\varepsilon^{ijk} e_i^{\phantom{i}a} \sigma_{a\alpha\dot\alpha}
\partial_j \bar\psi_k^{\phantom{k}\dot\alpha}\nonumber\\
&& -{1 \over 2} \varepsilon^{ijk}(\partial_i e_j^{\phantom{j}a})
\sigma_{a\alpha\dot\alpha} \bar\psi_k^{\phantom{k} \dot\alpha},
\label{eq:1}\\
\bar{S}_{\dot\alpha}&=&{i\over 2} P_{-\phantom{i}a}^{\phantom{-}i}
C^{\dot\beta \alpha}_{ji} \sigma^a_{\phantom{a}\alpha\dot\alpha}
\bar\pi^j_{\phantom{j}\dot\beta}+ \partial_i
  \bar\pi^i_{\phantom{i}\dot\alpha}\nonumber\\
&& -{1 \over 2} \varepsilon^{ijk} (\partial_i e_j^{\phantom{j} a} )
C^{\dot\beta \alpha}_{lk} \sigma_{a\alpha\dot\alpha}
\bar\pi^l_{\phantom{l}\dot\beta},
\label{eq:2}\nonumber\\
J_{\alpha\beta}&=&{1 \over 4}(\sigma^a \bar\sigma^b -
\sigma^b\bar\sigma^a )_\alpha^{\phantom{\alpha}\gamma}
\varepsilon_{\gamma\beta}\left(e_{ia}P_{-\phantom{i}b}^{\phantom{-}i}
+i\varepsilon^{ijk} e_{ia} \partial_j e_{kb} \right) \, ,
\\
J_{\dot\alpha \dot\beta}&=& -{1 \over 4}\varepsilon_{\dot\alpha \dot\gamma}
(\bar\sigma^a \sigma^b - \bar\sigma^b \sigma^a
)^{\dot\gamma}_{\phantom\gamma \dot\beta}
\left(e_{ia}P_{-\phantom{i}b}^{\phantom{-}i}
-i\varepsilon^{ijk} e_{ia} \partial_j e_{kb} \right)\nonumber\\
&&-{1 \over 2}
\left( \bar\pi^i_{\phantom{i}\dot\alpha}\varepsilon_{\dot\beta \dot\gamma}
\bar\psi_i^{\phantom{i}\dot\gamma}+ \bar\pi^i_{\phantom{i}\dot\beta}
\varepsilon_{\dot\alpha \dot\gamma} \bar\psi_i^{\phantom{i}\dot\gamma}
\right) \, . \nonumber
\end{eqnarray}
It remains to write down also the diffeomorphism and Hamiltonian
constraints $\bar{H}_{\alpha\dot{\alpha}}$. Here a crucial simplification
due to supersymmetry
occurs: On the classical level we have checked explicitely that
$\bar{H}_{\alpha\dot{\alpha}}(\bbox{x})\delta(\bbox{x}-\bbox{y})=
-2i\{S_\alpha(\bbox{x}),\bar{S}_{\dot{\alpha}}(\bbox{y})\}^*$
+ (terms proportional to $J_{\gamma\delta}$,
$\bar{J}_{\dot{\gamma}\dot{\delta}})$, where $\{\dots\}^*$ denotes the
usual
Dirac bracket for Grassmann-odd variables. Therefore, instead of
$\bar{H}_{\alpha\dot{\alpha}}$ we may equally well use generators
$H_{\alpha\dot{\alpha}}$ which quantum-mechanically
are defined via the supersymmetry generators.
\begin{equation}
\label{eq:3}
H_{\alpha\dot{\alpha}}(\bbox{x})\delta(\bbox{x}-\bbox{y})=-
  \frac{2}{\hbar}\left[S_\alpha(\bbox{x}),\bar{S}_{\dot{\alpha}}
   (\bbox{y})\right]_+
\end{equation}
This is the well-known supersymmetric square-root of gravity provided by
supergravity \cite{17}. Then a straightforward but, unfortunately, quite
tedious algebra which fortunately closely paralles the corresponding
calculation for the Bianchi models \cite{3,4} yields the following graded
generator algebra
\begin{eqnarray}
\label{eq:4}
 \left[S_\alpha(\bbox{x}),S_\beta(\bbox{y})\right]_+&=&0=
  \left[\bar{S}_{\dot{\alpha}}(\bbox{x}),
   \bar{S}_{\dot{\beta}}(\bbox{y})\right]_+\nonumber\\
\label{eq:5}
  \left[H_{\alpha\dot{\alpha}}(\bbox{x}),S_\beta(\bbox{y})\right]_-&=&
    i\hbar\delta(\bbox{x}-\bbox{y})(-\varepsilon_{\alpha\beta})
     {\bar{D}_{\ \dot{\alpha}}}^{\dot{\beta}\dot{\gamma}}(\bbox{x})
     \bar{J}_{\dot{\beta}\dot{\gamma}}(\bbox{x})\\
\label{eq:6}
   \left[H_{\alpha\dot{\alpha}}(\bbox{x}),\bar{S}_{\dot{\beta}}
    (\bbox{y})\right]_-&=&
    i\hbar\delta(\bbox{x}-\bbox{y})\varepsilon_{\dot{\alpha}\dot{\beta}}
     \bigg[D_\alpha^{\ \beta\gamma}(\bbox{x})
      J_{\beta\gamma}(\bbox{x})\nonumber\\
&&  +i\hbar\delta(0)
      \left({\bar{E}_\alpha}^{\dot{\gamma}\dot{\delta}}(\bbox{x})
       \bar{J}_{\dot{\gamma}\dot{\delta}}(\bbox{x})
        \right.\nonumber\\
&& -n^a(\bbox{x})h^{-1/2}(\bbox{x})\left.\sigma_{a\alpha\dot{\gamma}}
        \bar{S}^{\dot{\gamma}}(\bbox{x})\right)\bigg]\nonumber
\end{eqnarray}
and the usual commutators with $J_{\alpha\beta}$,
$\bar{J}_{\dot{\alpha}\dot{\beta}}$ which indeed we find to generate the
infinitesimal Lorentz transformation expected from the index-structure of
all the generators. The coefficients ${D_\alpha}^{\beta\gamma}$,
${\bar{E}_\alpha}^{\ \dot{\beta}\dot{\gamma}}$ are Grassmann-odd structure
functions. Their form is similar to the result for the Bianchi models
(see \cite{4}) and need not be given here as the explicit form is not
required in the following. We do note the divergent $\delta(0)$-factor
in the last of eqs.(\ref{eq:4}), however, which may hide an anomaly and
whose presence reduces this result to a formal
one. To go beyond this level one would have to introduce a regularization
first which renders $\delta(0)$ finite, then compute the commutator in
the regularized theory and check that the algebra still closes when
passing to the limit. This we
shall not attempt here. Fortunately, the last commutator in (\ref{eq:4})
is not
needed in our solution of the constraints, and the $\delta(0)$-term
therefore does not appear there. The only remaining (and, in pure gravity,
most difficult to evaluate) commutators $[H_{\alpha\dot{\alpha}}(x),
H_{\beta\dot{\beta}}(y)]$ follow immediately from
eqs.(\ref{eq:4}) by Jacobi-identities and are therefore easily obtained
here.
Again they evaluate to a linear combination of the generators
$S_\gamma$, $\bar{S}_{\dot{\gamma}}$, $J_{\gamma\delta}$,
$\bar{J}_{\dot{\gamma}\dot{\delta}}$, $H_{\gamma\dot{\gamma}}$
multiplied by structure functions from the left. Therefore we have
established that, formally, the graded generator algebra is closed on
the physical states, annihilated by all generators.

To determine a physical state explicitely we now follow the conjecture
of \cite{3} and make the ansatz
\begin{equation}
\label{eq:7}
 \Psi=\prod_{(\bbox{x})}S^\alpha(\bbox{x})S_\alpha(\bbox{x})
        g(\{{e_i}^a\})
\end{equation}
containing a formal product over all (suitably discretized) space-points
and, a yet undetermined bosonic functional $g$ independent of
${\bar{\psi}_i}^{\dot{\alpha}}$ satisfying $J_{\alpha\beta}g=0$,
$J_{\dot{\alpha}\dot{\beta}}g=0$. In the same
way as for the Bianchi models the ansatz (\ref{eq:7}) ensures that the
$S_\beta$-constraint and the $J_{\alpha\beta}$,
$\bar{J}_{\dot{\alpha}\dot{\beta}}$-constraints are
automatically satisfied. The $\bar{S}_{\dot{\beta}}$-constraint,
after using the generator algebra and the properties of $g$, is satisfied
if
\begin{equation}
\label{eq:8}
\bar{S}_{\dot{\alpha}}S_\alpha g(\{{e_i}^a\})=0\,.
\end{equation}
It is important to note that the operators ${e_p}^a
\bar{\sigma}_a^{\dot{\alpha}\alpha}\bar{S}_{\dot{\alpha}}S_\alpha$
and
$n^a\bar{\sigma}_a^{\dot{\alpha}\alpha}\bar{S}_{\dot{\alpha}}
S_\alpha$,are Lorentz-invariant, i.e. commute with $J_{\gamma\delta}$,
$\bar{J}_{\dot{\gamma}\dot{\delta}}$. For the Bianchi models a special
solution of eq.~(\ref{eq:8}) is found by solving $S_\alpha g=0$, and the
solution is in this case given by the restriction of the functional
$g_0(\{{e_i}^a\})=\exp[-\frac{1}{2\hbar}\int d^3x
\varepsilon^{ijk}{e_i}^a\partial_je_{ka}]$ to the appropriate spatially
homogeneous tetrad. But in the present {\it general} case,
$S_\alpha g=0$ has {\it no} solution, as shown in \cite{12,13,14}.
Remarkably, however, the more general equation (\ref{eq:8}) {\it does}
have solutions also in the present spatially inhomogeneous case, one of
which is, surprisingly, again given by the functional $g_0$. However,
while $\bar{J}_{\dot{\alpha}\dot{\beta}}g_0=0$ is satisfied, one
checks that $J_{\alpha\beta}g_0\neq 0$. A fully Lorentz-invariant
amplitude $g$ is obtained from $g_0$ only after
explicit symmetrization with respect to the transformations generated
by the three generators $J_{\alpha\beta}$. Thus
$g(\{{e_i}^a\})=\int D\mu[\omega]\exp
(i\omega^{\alpha\beta}J_{\alpha\beta})g_0(\{{e_i}^a\})$. Here
$D\mu[\omega]$ is chosen as the formal direct product of the Haar
measure of the $SU(2)$-rotation matrices ${\Omega_\alpha}^{\ \beta}=
\left[\exp(i{\omega_.}^{\ .}\right]_{\alpha}^{\ \beta}$ with
${\omega_2}^{\ 2}=({\omega_1}^1)^*$, ${\omega_2}^1=-({\omega_1}^2)^*$.
The symmetrized amplitude $g$ is still a solution of
(\ref{eq:8}) because, as was already noted, the operator on the left
is proportional to
Lorentz-invariant operators, which therefore commute with the symmetrizing
rotations. Rewriting the infinite product in (\ref{eq:7}) as a
Grassmannian path-integral over a Grassmann field $\varepsilon^\alpha
(\bbox{x})$,
applying the factors $S^\alpha(\bbox{x})$ explicitely on the functional
$g$, and using the identity
\begin{eqnarray*}
&& \exp(i\omega^{\alpha\beta}J_{\alpha\beta})
g_0(\{{e_i}^a\})=[\exp(i\omega^{\alpha\beta}J_{\alpha\beta})
g_0^2(\{{e_i}^a\})\\
&&\mbox{\hspace{3cm}}
\exp(-i\omega^{\alpha\beta}J_{\alpha\beta})]
[g_0(\{{e_i}^a\})]^{-1}
\end{eqnarray*}
satisfied by $g_0$
we obtain our exact result for the physical state $\Psi$ in the final form
\begin{eqnarray}
\label{eq:9}
&&\Psi(\{h_{ij},{\bar{\psi}_i}^{\ \dot{\alpha}}\})=
    \int D[\varepsilon^1]D[\varepsilon^2]
     \Bigg\{\exp\bigg[-\int d^3x\varepsilon^{ijk}\nonumber\\
&&\bigg[
      \varepsilon^\alpha(\bbox{x})\partial_j\sigma^a_{\alpha\dot{\alpha}}
       {\bar{\psi}_k}^{\ \dot{\alpha}}(\bbox{x})e_{ia}(\bbox{x})
         +\frac{1}{2\hbar}{e_i}^a(\bbox{x})\partial_je_{ka}(\bbox{x})
          \bigg]\bigg]\nonumber\\
&& \qquad \int D\mu[\omega]
     \exp
     \Big[\int d^3x\varepsilon^{ijk}{\Omega_\gamma}^\alpha(\bbox{x})
      (\partial_j{\Omega^\gamma}_\beta(\bbox{x}))
       \nonumber\\
&& \mbox{\hspace{1cm}}
         \sigma_{a\alpha\dot{\alpha}}{e_i}^a(\bbox{x})
         \Big(\varepsilon^\beta(\bbox{x}){\bar{\psi}_k}^{\dot{\alpha}}
          (\bbox{x})+\frac{1}{2\hbar}
           {\bar{\sigma}_b}^{\ \dot{\alpha}\beta}
          {e_k}^b(\bbox{x})\Big)\Big]\Bigg\}\,.
\end{eqnarray}
That the right-hand side of eq.~(\ref{eq:9}) is, indeed, a function of
the 3-metric $h_{ij}$ follows from the invariance under the Lorentz
generators, which makes one free, without restriction of generality, to
choose ${e_i}^a$ in the argument in the form ${e_i}^0=0$,
${e_i}^{\hat{a}}=q_{i\hat{a}}$, $\hat{a}=1,2,3$ where $q_{ij}=q_{ji}$
denotes the positive definite, symmetric matrix square-root of $h_{ij}=
\sum_{k=1}^3q_{ik}q_{jk}$ which is uniquely determined by $h_{ij}$.

Let us now briefly discuss our result. First we note that it seems to be
close to but not identical with a result recently obtained by Matschull
\cite{18} along quite different lines using a new representation
somewhere in-between the metric representation employed here and the
Ashketar representation. Like our final result Matschull's also
contains functional integrals over a spatial 2-component Grassmann field
and a spatial field of $SU(2)$-rotation matrices. Among the
differences with our result the most obvious one is the
sign in the exponent of the unsymmetrized bosonic amplitude
$g_0(\{{e_i}^a\})$. This sign would
also be changed in our calculation by using a different operator
ordering, but actually we have no freedom in the choice of this sign,
if we wish to
reproduce correctly the normalizable amplitude \cite{19}
$\sim\exp[-\frac{V}{2\hbar}m^{pq}h_{pq}]$ of the spatially homogeneous
Bianchi-models \cite{3,4,6,7,8a,8b,9}. The fact that our amplitude
$g$ reduces to this form in the spatially homogeneous case indicates that
(\ref{eq:9}) should be interpreted as a worm-hole state. As the state
obtained in \cite{18} does not similarly fall off for large spatially
homogenous 3-geometries, it cannot be interpreted in this way. Apart
from this and the already
mentioned results for the Bianchi class A models there seems to be no
analytical result to compare with. It appears likely, however, and would
indeed be very interesting to verify, that the semiclassical
wave-functional obtained for $N=1$ supergravity with non-vanishing
cosmological constant
$\lambda$ in the Ashketar representation \cite{20} reduces to the present
result in the (quite nontrivial) limit $\lambda\to 0$. At least for the
Bianchi class A models this happens to be the case \cite{21}.

The bosonic amplitude $g$ is somewhat reminiscent of
the ground-state functional of quantum electrodynamics when the latter is
written in terms of the transversal part of the vector potential. To
obtain an exact solution of this kind for the Wheeler DeWitt equation of
quantum gravity has been an
outstanding goal for a long time after it was formulated by Wheeler
\cite{22}. This (seemingly modest) goal has so far eluded its attainment
in the case of pure gravity. Matschull \cite{18} recently was able to
construct a solution of the quantum gravity constraints in a new
representation; however he also found that the same construction gives
rise to a Lorentz non-invariant quantum correction in the Wheeler DeWitt
equation.
It is therefore quite remarkable (and another instance of the
simplifying nature of supersymmetry) that the same obstruction does
not occur in the case of supergravity, where the Wheeler DeWitt operator
is replaced by $n^a{\bar{\sigma}_a}^{\ \alpha\dot{\alpha}}
\bar{S}_{\dot{\alpha}}S_\alpha$ in which the anomalous
term does not occur, and where therefore Wheeler's goal is attained by
eq.~(\ref{eq:9}). In view of the highly nonlinear form of gravity and
supergravity it is very surprising that this result (apart from the
Grassmannian component) turns out to be so similar to the Gaussian form
expected
for the ground state of a free field like the free electromagnetic field.
Perhaps there is some hope, therefore, that other exact quantum states
corresponding to gravitons (in the same sector) or pairs of
gravitinos (in sectors differing by an even fermion number) may be found.
To see whether this hope is justified, and also to find the explicit form
of a Hartle-Hawking state (see \cite{4,26} for the spatially homogeneous
case), remain interesting open problems for future work.

This work has been supported by the Deutsche Forschungsgemeinschaft through
the Sonderforschungsbereich 237 ``Unordnung und gro{\ss}e Fluktuationen''.
One of us (A.~Csord\'as) would like to acknowledge additional support by
The Hungarian National Scientific Research Foundation under Grant number
F4472.


\begin{thebibliography}{99}
\bibitem{18a} The ordering in the generators $S_\alpha$,
              $\bar{S}_{\dot{\alpha}}$, $H_{\alpha\dot{\alpha}}$ are
              closely related by supersymmetry and are fixed by the one
              free choice of ordering of
              $P_{-\phantom{i}a}^{\phantom{-}i}$ and
              $C_{ji}^{\dot{\beta}\alpha}$ in $\bar{S}_{\dot{\alpha}}$.
\bibitem{19} Here ${C^p}_{rs}=m^{pq}\varepsilon_{qrs}$ are the
             structure constants of the Bianchi models in class
             A.
\end{thebibliography}
\end{document}